\documentstyle[12pt,epsfig]{article}
\begin{document}
\noindent
\begin{center}
{\Large {\bf The Need for Curriculum Updating in Physics Education \\
}} \vspace{2cm}
 ${\bf Yousef~Bisabr}$\footnote{e-mail:~y-bisabr@sru.ac.ir}\\
\vspace{.5cm} {\small{Department of Physics, Shahid Rajaee Teacher
Training University,
Lavizan, Tehran 16788, Iran.}}\\
\end{center}
\vspace{1cm}
\begin{abstract}
We argue that most advances in science and technology during the past and the present centuries rely on modern physics concepts developed in the early $20$th century. The reliance is so profound that any improvement of literacy of science and technology falters in the absence of those concepts in physics education. Despite this important role, there are some remarks in literature concerning that modern topics have gone unnoticed in physics textbooks \cite{cht}. We discuss confirmation of this claim and also argue that physics education without reference to the difficulties of classical concepts and the need for paradigm change can lead to serious misconceptions and gives an incorrect image of science.
\end{abstract}
\vspace{.7cm}

\vspace{1cm}
\section{Introduction}
Many innovative and stunning advances of science and technology over the past and the present centuries profoundly affect our lives. In particular, advances in satellite communications, computers and smartphones technologies  have made it possible to develop an internet-based life activities.  Due to this huge and fast scientific developments, it is not so hard to predict that even more fascinating capabilities such as quantum computing, space travels, virtual reality, nano-medicines and many others will enter our lives in near future.
Most of these advances rely on two revolutionary physical theories developed during the first quarter of 20th century, namely the theory of relativity and the quantum theory. These theories have so profound effects on our view of the Universe that physics beyond them is named by the term \emph{modern physics}.\\
 The theory of relativity consists of the special and the general theories. The special theory was introduced by Einstein in 1905. In this theory Einstein challenged the absoluteness of space and time, which had been assumed in classical physics, and brought these two concepts into
a unique entity known as spacetime. Ten years later in 1915, Einstein introduced the general theory of relativity
as a generalization of the special theory. In the general theory, Einstein provided a refinement of Newton's
gravitational theory and interpreted gravity as a geometric property of spacetime, the curvature of spacetime in a region is connected to distribution of matter in that
region by a set of dynamical field equations, the so-called Einstein field equations. General relativity has several important astrophysical implications such as deviations from the classical precession of Mercury's perihelion, bending of light, existence of black holes and gravitational waves. Some of these predictions were verified experimentally soon after introducing the theory and the others verified later.\\
On the other hand, quantum theory was started by Planck's explanation of blackbody radiation in 1900 and Einstein's explanation of the photoelectric effect in 1905. Then the theory was developed by several well-known physicists such as, Bohr, de Broglie, Heisenberg, Schrodinger and many others. In this theory the determinism of classical physics is replaced by indeterminism induced by the wave-particle duality and the uncertainty principle.\\
Modern physics affects our lives in two ways. First, it has revolutionized our philosophical and scientific world view. For instance, in classical physics all observers, equipped with standard clocks and
 rods, agree with outcomes of measuring space and time intervals. This is not true in the modern viewpoint in which those measurements are observer-dependent.
  On the other hand, outcomes of quantum measurements are no longer deterministic and are represented by probabilities and statistical quantities.\\
Second, modern physics has a very strong influence on daily life by introducing various technological innovations. Some examples are the global positioning system (GPS) which is nowadays used for navigating aircrafts, ships and vehicles; fission-based nuclear energy dating back to the last mid-century and also fusion-based nuclear energy that seems to be accessible in near future; marvelous tools for studying human bodies and treating diseases and improving health; advanced electronic devices and communicating systems which all have their origins in semiconductor chips and the list goes on.\\
Despite the large dependency of our lives on these modern technologies, one can hardly find anyone who knows anything about their scientific origins \cite{sagan}. This gap between technological advances and scientific/technological literacy of individuals in a society can potentially slow down scientific developments both in modern and  developing countries. This gap needs to be filled by education policies and science education communities if they wish to increase effective participation of citizens and prepare them for living in the new century.\\
In this work, we deal with this issue and argue that one of the ways to address the problem is to enter gradually basic concepts of modern physics in high schools physics curricula. In section 2, we briefly describe how some popular technologies, that we are using in our everyday lives, are related to the revolutionary ideas within modern physics. In section 3, we discuss some reasons for entering modern physics in high schools. However, the process of updating physics curricula should be managed cautiously since there are also obstacles on this way which can potentially make serious difficulties. In section 4, we outline our conclusions.
~~~~~~~~~~~~~~~~~~~~~~~~~~~~~~~~~~~~~~~~~~~~~~~~~~~~~~~~~~~~~~~~~~~~~~~~~~~~~~~~~~~~~~~~~~~~~~~~~~~~~~~~~~~~~~~~~~~~~~~~~~
\section{Modern physics in our daily lives}
Society's reliance on  modern technologies manifests the key role played by modern physics in today's life. Many modern devices or possibilities that we use in our everyday life would not have been available without the revolutionary ideas proposed in the early of $20$th century. In this section, we briefly review some of these ideas and their interrelations to modern technologies.
\subsection{The theory of relativity}
Nuclear energy is one of the consequences of the theory of relativity.
Before $20$th century, all energy resources in the world were limited to fossil resources. By developing the special theory of relativity, Einstein introduced the mass-energy equivalence formula $E=mc^2$ \cite{e} defining a relationship between the mass $m$ and the energy $E$ of a body in its rest frame\footnote{The rest frame is the frame of reference where the body is at rest.}. In this relation the square of the speed of light is an enormous number, and therefore a small amount of the rest mass is associated with a tremendous amount of energy. This then opened a new window for energy production compared to fossil resources.\\
Nuclear energy power can be based on two kinds of nuclear reactions, fission and fusion. In the fission process the heavy nucleus of an atom, such as that of uranium or plutonium, breaks up into two or more lighter nuclei. The process may take place spontaneously in some cases or may be induced by excitation of the nucleus with a variety of particles. Nuclear fission of heavy elements was discovered in $1938$ and soon after that discovery, different energy power plants based on this process were used around the world. \\
Nuclear fusion is the process by which light atomic nuclei combine to form a heavier one while releasing huge amounts of energy.  The process powers active or main sequence stars, such as the sun, in which nuclei collide with each other at extremely high temperatures. The high temperature provides nuclei with enough energy to overcome their mutual electrical repulsion. Fusion power offers the prospect of an almost inexhaustible source of energy for future generations. Although, fusion-based energy production presents so far unresolved engineering challenges, it is an important candidate as a clean energy resource in near future.\\
Another consequence of the theory of relativity is the global positioning system (GPS).
The GPS project was firstly started with military purposes in $1970$s and then civilian use was allowed from the 1980s. Nowadays, GPS is wildly used for navigating aircrafts, shipping,
in private and commercial vehicles, and for urban navigation. It consists of a constellation of twenty-four satellites
in six orbital planes, in each of which reside four satellites, in high (around $20000$ km
radius) orbits all with a period of twelve hours. Each satellite carries an accurate record of its position and time and transmits that data to the receiver. The satellites carry very stable atomic clocks that are synchronized with one another and with ground clocks. Any drift from time maintained on the ground is corrected daily.\\
 This network of satellites is designed so that any point on the earth is covered by at least four satellites. GPS receivers on the earth (in flight or on a ship or on the ground) detect signals that emitted by
these satellites. The position of receivers  are then determined by means of triangulation.
The distance between
the satellites and a GPS receiver is the time difference between emission
and detection of this signal multiplied by the speed of light. For calculating the position of the receiver, it is therefore necessary to keep track time to high accuracy and this is where relativistic effects come into play.
Due to relative motion of the clocks, there is a special relativistic time dilation. Moreover, a more precise calculation takes into account
the fact that the frames are not inertial since they are rotating. There is also a general relativistic
effect since the clocks on the earth and those in the satellites are in different heights in the earth's gravitational potential. This second effect generally deals with slowing down of clocks in a gravitational potential well
with respect to clocks far away. The accumulated time error due to neglecting the two aforementioned relativistic effects is $3.8\times 10^{-5}$ seconds which corresponds to a position error $11.4 km$ \cite{faraoni}. The GPS system automatically corrects for both relativistic effects and without these corrections this system would become completely useless.
The arguments show how the seemingly abstract theory of relativity
 have become essential to the functioning of modern life.
 \subsection{Quantum physics}
Quantum physics plays a fundamental role in the development of semiconductors and transistors in the $1940$s. Many modern electronic devices that we use in our lives are designed using these objects. There are numerous examples including computers and laptops, smartphones, TV and video sets and satellite receivers.\\
Moreover, quantum physics also has an important role in medical sciences. There are several devices used for medical diagnosis which are designed and worked based on quantum physics. For example, X-ray imaging and radiation therapy have been in use for over a century, positron emission tomography (PET) and magnetic resonance imaging (MRI). Besides developments of these devices, there are several scientific areas which have benefited significantly from the introduction of nanotechnology. An example is the nanomedicines which are medical applications of nanotechnology and ranges from applications of nanomaterials to nanoelectronic biosensors \cite{nano}.\\
~~~~~~~~~~~~~~~~~~~~~~~~~~~~~~~~~~~~~~~~~~~~~~~~~~~~~~~~~~~~~~~~~~~~~~~~~~~~~~~~~~~~~~~~~~~~~~~~~~~~~~~~~~~~~~~~~~~~~~~~~~~~~~~~
\section{Modern physics in physics education}
Despite the fact that relativity theory and quantum physics are over a century old, introductory physics classes essentially ignore them \cite{aub}.
The regular high school physics textbooks and even non-major undergraduate universities physics courses usually cotain concepts which have been
elaborated during $18$th and $19$th centuries. Science education is aimed at
increasing students’ scientific literacy and their ability to make decisions in
matters that impact their daily lives and their future as adults.
However, it is not possible to reach this aim by ignoring scientific revolutionary ideas introduced and developed during $20$th century.\\
In the last few decades, there have been consolidation of an international consensus concerning the need to include various modern physics concepts in the high school physics curricula \cite{a} \cite{b} \cite{I}.
Some of the reasons for this demand are the followings:\\\\
1) As pointed out in the previous section, most technological innovations and advances are based on the theories of modern physics introduced and developed in the last century. Since technological literacy is the ability of a person to use, manage and understand technology, one can not expect that such an ability can be achieved without an enough realization of basic concepts of modern physics.\\\\
2) During the last few decades, the progresses in satellite communications have been led to a very large accessibility to internet and social media networks all around the world. Due to these new possibilities, everyone can easily access to the latest scientific news, scientific TV programmes and science-fiction films, series and animations. This makes scientific idioms such as black holes, wormholes, time travel or parallel words enter public conversations. Scientific idioms that have been already used only by scientists are now popular among non-specialist people and pupils or even kids. This may be regarded as a new social phenomenon which is specific to this century. It will force education policies to be adapted with this new condition. Science teachers and scientific curricula, as the most important parts of an education system, should be prepared for answering students' questions about these specialized subject matters.\\\\
3) The classical physics concepts such as the conservation of mass and the conservation of energy, the absoluteness of space and time intervals and the concept of gravitational force in Newtonian gravity have central roles in introductory physics curricula. These are just approximations of the modern concepts and may give students an incorrect and simplistic image of science.\\\\
4) The over-emphasis on classical concepts leads to an ignorance of the crisis existence and the need for paradigm shifts \cite{b}. To enhance students' interest, physics should be taught as a growing subject. It is also important to give students illustrations that indicate problems of the frontiers and then the need to change the paradigm.
~~~~~~~~~~~~~~~~~~~~~~~~~~~~~~~~~~~~~~~~~~~~~~~~~~~~~~~~~~~~~~~~~~~~~~~~~~~~~~~~~~~~~~~~~~~~~~~~~~~~~~~~~~~~~~~~
\section{Discussions}
Science has an increasingly important role in our lives. It improves our quality of living and gives a better understanding of the world around us. Despite this critical role, most individuals have low realizations of science and technology. In fact, there is a gap between rapid advances of modern societies and scientific literacy of individuals. This gap should be filled by education systems and appropriate education policies.\\
In this work, we have argued that most technological advances and innovations are based on the physical theories introduced and developed during $20$th century. In particular, we dealt with the interrelation between these advances and relativity theory and quantum physics. Despite the important role of modern physics concepts, they have almost no places in physics curricula in high schools or even in non-major university courses and most physics topics there contain the classical conceptions developed during $18$th and $19$th centuries. This traditional method in physics education seems not to be consistent with many rapid technological progresses taken place during the last few decades. In particular, the progresses in satellite communications, internet and social media have led to promotion and popularization of modern physics concepts and idioms. This necessitates using a modern method in physics education in which the modern physics concepts enter physics curricula and textbooks.\\
However, there are obstacles on this way. The first is that the theories of modern physics are usually formulated in highly mathematical formalisms which are not familiar to school students and are just suitable for sufficiently skillful and gifted students \cite{c1}. It should be noted that classical physics theories, such as Newtonian gravity, can also be extremely mathematically complex but those aspects are not taught
at high schools. The modern topics should be taught with an emphasis on the observational points and without recourse to complex mathematical concepts and formulas \cite{I}. Moreover, basic concepts
of modern physics such as relativity of simultaneity, wave-particle duality and so on are too difficult to conceptualize and comprehend without
a methodical introduction \cite{c2}. In other words, theories of modern physics are based on laws and principles that seems to be less intuitive than those of classical physics. For instance, measurements of quantities such as lengths and durations are observer-dependent in the theory of relativity or matter has a wavy nature in quantum physics. Although supported by strong experimental evidences\footnote{In the former case, the special and general relativistic time dilation in GPS described above and in the latter case, the electron diffraction experiment are typical examples.}, they are not consistent with classical intuition of students. As mentioned above, the use of experimental evidences and appropriate analogies \cite{I} \cite{II} improve teaching modern physics at high schools.

~~~~~~~~~~~~~~~~~~~~~~~~~~~~~~~~~~~~~~~~~~~~~~~~~~~~~~~~~~~~~~~~~~~~~~~~~~~~~~~~~~~~~~~~~~~~~~~~~~~~~~~~~~~~~~~~~~~~~~~~~~~~~

\end{document}